# RESULTS FROM A PROTOTYPE PERMANENT MAGNET DIPOLE-QUADRUPOLE HYBRID FOR THE PEP-II B-FACTORY*

M. Sullivan, G. Bowden, S. Ecklund, D. Jensen, M. Nordby, A. Ringwall, Z. Wolf
Stanford Linear Accelerator Center, Stanford University, Stanford, CA 94039 USA

*Abstract*

We describe the construction of a prototype hybrid permanent magnet dipole and quadrupole. The magnet consists of two concentric rings of $Sm_2Co_{17}$ magnetic material 5 cm in length. The outer ring is made of 16 uniformly magnetized blocks assembled as a Halbach dipole and the inner ring has 32 blocks oriented in a similar fashion so as to generate a quadrupole field. The resultant superimposed field is an offset quadrupole field which allows us to center the field on the high-energy beam in the interaction region of the PEP-II B-factory. The dipole blocks are glued to the inside surface of an outer support collar and the quadrupole blocks are held in a fixture that allows radial adjustment of the blocks prior to potting the entire assembly with epoxy. An extensive computer model of the magnet has been made and from this model we developed a tuning algorithm that allowed us to greatly reduce the n=3–17 harmonics of the magnet.

## 1 INTRODUCTION

Obtaining high luminosity from modern colliding beam accelerators generally requires magnetic components to be within 2 m of the collision point. Accelerator elements this close to the interaction point (IP) are invariably inside a detector solenoidal magnetic field. This restricts the choice of technologies for these elements to either superconducting or permanent magnet (PM).

For PEP-II, the *B*-factory being designed and built by a collaboration from SLAC, LBNL, and LLNL[1,2], the 9 and 3.1 GeV beams collide head-on and then are horizontally separated by a bend magnet (B1) positioned between 0.2 and 0.7 m from the IP. Separation continues when the two beams travel through the next magnetic element Q1, a horizontally defocusing quadrupole extending between 0.9 and 2.1 m from the IP. The separation of the beams is maximized if the high-energy beam (HEB) travels through the magnetic center of Q1. Both B1 and Q1 are inside the magnetic field of the detector and are therefore made from PM material.

## 2 Q1 MAGNET DESIGN AND BLOCK SPECIFICATIONS

The Q1 magnet is the vertically focusing quadrupole of the final focus doublet for the low-energy beam (LEB)

*Work supported by U.S. Department of Energy, under contract number DE-AC03-76SF00515.

with a gradient of 106 kG/m. The field quality of this magnet must be very good; we are asking for the strength of each higher harmonic to be less than $1\times10^{-4}$ of the quadrupole strength at a radius that contains the beam-stay-clear (BSC) apertures of both the LEB and the HEB. This requirement is especially demanding at the back of the magnet where the beams are farthest apart (a circle encompassing both BSCs has a radius of 5.9 cm at this position). The BSC is defined as $15\sigma+2$ mm where $\sigma$ is the calculated beam size using uncoupled emittance for x and fully coupled emittance for y[3].

In order to minimize the space taken up by the Q1 magnet, we decided to horizontally shift the magnetic center of Q1 about 2 cm by superimposing a 2.1 kG dipole field over the quadrupole field. This positions the magnetic center of Q1 over the HEB orbit and keeps both beams relatively close to the mechanical center of the magnet thus minimizing harmonic feed-down effects.

Figure 1 shows the positions and magnetization angles of the blocks. Q1 is made up of uniformly magnetized blocks of $Sm_2Co_{17}$ PM material. The magnet is divided into 5 cm thick slices with each slice containing 32 blocks oriented and magnetized according to the method described by Halbach[4] to make a quadrupole field. Outside of the ring of 32 blocks is another ring of 16 blocks oriented and magnetized to make a horizontal dipole field. Because of the limited space for this strong magnet, the blocks are held in place with epoxy. The inside radius of the magnetic material is 7.4 cm.

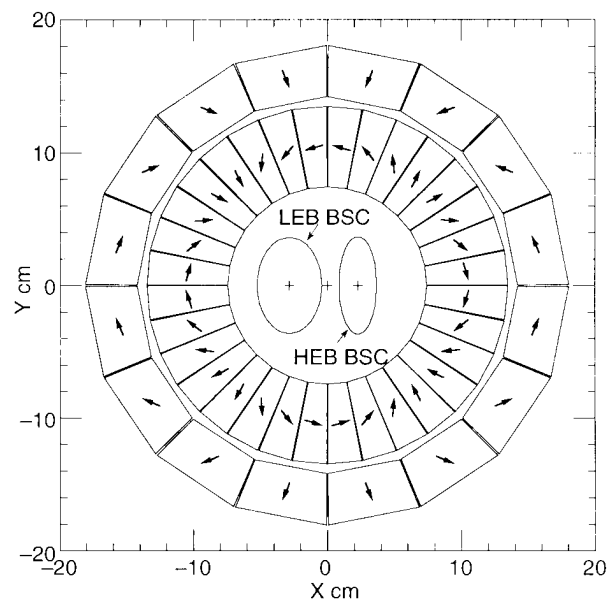

Figure 1. Block dimensions and magnetization angles of the blocks that make the field in the Q1 magnet. The BSC ellipses are at the end farthest away from the IP (z=2.1 m).

A computer model of the magnetic blocks has been made which computes the magnetic field for each magnetic block according to the Halbach formulas and sums up the individual block fields to get the total magnetic field of a slice. This program structure allows one to model inaccuracies in block placement, dimensions and magnetization. The introduction of errors in the block parameters allows us to determine the expected harmonic content of a magnet slice. The program was used to help define the tolerance ranges for the magnetic blocks.

The unallowed higher harmonics in a Halbach style magnet stem from the block inaccuracies mentioned above. There are also allowed harmonics at n=m+N where m is the number of blocks in the magnet (m=32 for the quadrupole and 16 for the dipole in this case) and N=1 for a dipole and 2 for a quadrupole. To minimize the unallowed higher harmonics, we set fairly tight tolerances on the manufactured blocks.

The magnet blocks were machined from a Shin-Etsu material (R26HS) with a very high intrinsic coercive force (in excess of 18 kOe). Each magnetized block was thermally stabilized in an open circuit condition at 150°C. The magnetization strength of all blocks had to be within 2% of the average value with no block below 10.4 kG. The magnetization angle had to be within 2° of the correct value. The block dimensions were specified assuming a nominal block strength of 10.4 kG with tolerances on the block dimensions of ±200 µms. We attempted to position the dipole blocks to within ±200 µms radially and ±400 µms in azimuth of the correct location.

To verify the quality of the blocks, the dipole moment and magnetization angle of each block were measured using a rotating table positioned in the center of a Helmholtz coil. The magnetic moment measurement precision is 0.1% for the magnitude and 0.1° for the direction.

## 3 MAGNET ASSEMBLY

The first step in the assembly of a magnetic slice is to glue aluminum tabs to the inside and outside radial surfaces of the magnetic blocks. The adhesive is a fast-cure, high-temperature, modified acrylic (Loctite®325). Part of these tabs protrude above the block and allow for a fixture to bolt to these extensions holding the block rigidly. The fixture, with magnetic block, is then mounted on an xyz table for insertion into an aluminum collar mounted on a rotating table. The collar has been faceted on the inside diameter using a wire EDM to accommodate the 16 dipole blocks. The dipole blocks are glued to the inside of the collar with the Loctite®325 adhesive.

Once the dipole blocks are mounted in the collar, the quadrupole blocks can be placed into position. The 32 quadrupole blocks are prepared in the same manner as the dipole blocks. In this case, a brass block bolts to the two glue-tabs and this assembly is mounted into a slot in a support arm that positions the quadrupole block in the bore of the dipole magnet. The brass block is secured in the slot in the arm with a shoulder bolt that can be positioned radially with set screws. This mechanism allows us to independently adjust the radial position of each quadrupole block prior to epoxying the entire assembly. Once the slice is epoxied, the fixtures are removed and the tops of the glue-tabs machined off.

## 4 MAGNET TUNING AND HARMONIC MEASUREMENTS

The harmonics of a magnet slice are measured with a rotating bucking coil. The rotating assembly consists of three coils. Two coils are bucking coils, one designed to buck out the dipole field and the other to buck out the quadrupole field. The third coil measures the higher harmonics. The coil is 100 cm long with an effective radius of 6.35 cm and is mounted vertically on a set of rails so that it can be easily and quickly lifted out of the bore of the magnet. This gives us access to the bore to make adjustments to the quadrupole blocks. The signal from the rotating coil is fed to an integrator and then read out by a PC. The system has a sensitivity of 0.003% of the quadrupole signal for the higher harmonics.

The program that models the block tolerances can also simulate a perfect magnet slice. One can then introduce a sinusoidal variation (as a function of azimuth) of the radial position of the quadrupole blocks. This results in the generation of a pure sextupole and dipole harmonic. Introducing a radial block motion with a $\sin 2\theta$ azimuthal dependence produces a pure octupole (n=4) harmonic. The amplitude of the radial motion is directly proportional to the strength of the harmonic. The skew and normal harmonics can be independently generated by adjusting the phase of the radial block motion (e.g. $\sin\theta$ and $\cos\theta$). Inverting this process, one can compute the radial block motion needed to cancel measured harmonics in a magnet slice; the final block motion being the sum of the motions computed for each harmonic. This procedure works for a total of 32 normal and skew components; harmonics n=3 through 17, both skew and normal, and n=18, skew component only. One can also adjust the quadrupole normal harmonic by moving all of the blocks radially in or out. The normal component of n=18 cannot be changed because this harmonic produces a block motion function that matches the block segmentation in azimuth and therefore all blocks are located where the radial oscillation function crosses zero.

According to the computer model algorithm, the largest block motion needed to correct the harmonics prior to tuning was about 1 mm. The initial block adjustments were made by observing the set screw motion (counting bolt head flats). After the block displacement was reduced to less than about 250 µms, we used a capacitive readout sensor to monitor the motion of the blocks. This sensor is attached to a fixture that can be directed to the face of any one of the quadrupole blocks and can measure a difference in block position of less than 2 µms. Using the sensor, we were able to reduce the maximum block displacement to 28 µms (five blocks had displacements greater than 20 µms, the rest had values that were less than 15 µms). The harmonics of the magnet before and after tuning are shown in Figure 2. It should be noted that this procedure will tune out harmonics from any source. In particular, some of the harmonics from the dipole blocks are present in the magnet aperture and these harmonics have also been tuned out by adjusting the positions of the quadrupole blocks.

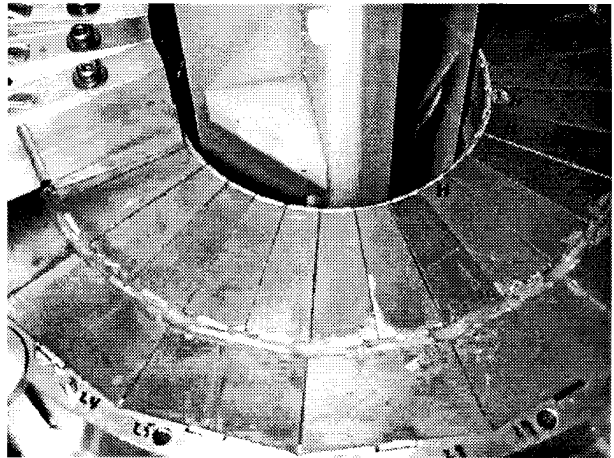

Figure 3. The prototype slice with the harmonic measuring coil in place. The Al tabs have been machined off. Some of the quadrupole block support arms are shown on the left side of the picture.

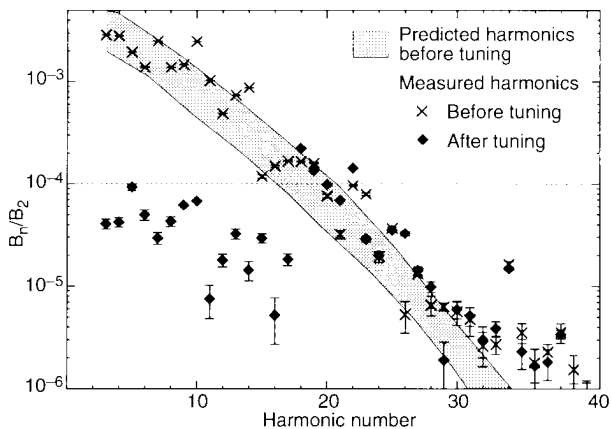

Figure 2. The harmonic content of the prototype magnet before and after the quadrupole blocks are radially adjusted. The quadrupole and dipole harmonics are off-scale and hence not shown. The inner radius of the magnetic material is 7.4 cm. These harmonics are shown at a radius of 5.9 cm. The gray area is the region for the harmonics predicted by the computer model before tuning.

After tuning, the entire magnet (quadrupole and dipole) was potted with a slow-curing, low-viscosity, casting epoxy (Epic®R1055/H5039, a highly cross-linked thermoset which contains a silica filler). We saw no change in the harmonic content of the slice after the epoxy had set and the fixtures had been removed. Figure 3 is a picture of the potted magnet. Some small changes (about $3 \times 10^{-5}$ at a radius of 6.35 cm) in some of the harmonic values were observed after about one month which we attribute to magnetic aging, but the following month saw no further change in the harmonics of the slice.

## 5 SUMMARY

We have constructed a 5 cm thick prototype permanent magnet quadrupole with a superimposed dipole field. The magnet is made of two concentric rings of uniformly magnetized blocks–16 for the dipole field and 32 for the quadrupole field–magnetized and oriented according to the Halbach formulas. A computer model of the magnet has been made that accurately predicts the harmonic quality of the magnet. The magnetic blocks are held in place with epoxy due to limited radial space for the magnet. Prior to potting, the quadrupole blocks are held in a fixture that allows the blocks to be radially adjusted. This ability is used to lower the harmonics (n=3–17) of the magnet to below $1 \times 10^{-4}$ of the quadrupole strength at a radius of 5.9 cm (80% of the magnet aperture).

## 6 ACKNOWLEDGMENTS

We would like to thank the machine shop and support staff at SLAC for their assistance in this study. We also thank the designers and technicians who helped us, especially Joe Stieber, who was the lead designer and Bruce Anderson, who was instrumental in helping us with the magnet assembly.